\documentclass[a4paper,fleqn]{cas-dc}
\usepackage[numbers]{natbib}
\usepackage{caption}
\usepackage{url}
\usepackage{xcolor}
\usepackage{color,soul}
\usepackage[textwidth=  15mm]{todonotes}
\usepackage{environ} 
\usepackage{etoolbox}

\graphicspath{{./images/}}

\newtoggle{doShowFigures}        
\NewEnviron{maybeShowFigures}{%
  \iftoggle{doShowFigures}{\BODY}{}%
}
\toggletrue{doShowFigures}
\begin{document}
\let\WriteBookmarks\relax
\def\floatpagepagefraction{1}
\def\textpagefraction{.001}

\shorttitle{NapierOne: A Modern Mixed File Data Set Alternative to Govdocs1}
\shortauthors{S.R.Davies,R.Macfarlane,W.J.Buchanan}

\title [mode = title]{NapierOne: A Modern Mixed File Data Set Alternative to Govdocs1}                      
\author[1]{Simon R. Davies}[type=editor,
                        auid=000,bioid=1,
                        orcid=0000-0001-9377-4539,]
\cormark[1]
\ead{s.davies@napier.ac.uk}
\address[1]{School of Computing, Edinburgh Napier University, Edinburgh, UK.}

\author[1]{Richard Macfarlane}[orcid=0000-0002-5325-2872]

\author[1]{William J. Buchanan}[orcid=0000-0003-0809-3523]

\begin{abstract}
 It was found when reviewing the ransomware detection research literature that almost no proposal provided enough detail on how the test data set was created, or sufficient description of its actual content, to allow it to be recreated by other researchers interested in reconstructing their environment and validating the research results. A modern cybersecurity mixed file data set called \emph{NapierOne} 
 is presented, primarily aimed at, but not limited to, ransomware detection and forensic analysis research.  \emph{NapierOne} was designed to address this deficiency in reproducibility and improve consistency by facilitating research replication and repeatability. The methodology used in the creation of this data set is also described in detail. 
 The data set was inspired by the Govdocs1
 data set and it is intended that \emph{NapierOne} be used as a complement to this original data set. 
 
An investigation was performed with the goal of determining the common files types currently in use. No specific research was found that explicitly provided this information, so an alternative consensus approach was employed. This involved combining the findings from multiple sources of file type usage into an overall ranked list. After which 5,000 real-world example files were gathered, and a specific data subset created, for each of the common file types identified. In some circumstances, multiple data subsets were created for a specific file type, each subset representing a specific characteristic for that file type. For example, there are multiple data subsets for the ZIP file type with each subset containing examples of a specific compression method. Ransomware execution tends to produce files that have high entropy, so examples of file types that naturally have this attribute are also present. The resulting entire data set comprises of a 100 separate data subsets divided between 44 distinct file types, resulting in almost 500,000 unique files in total.
A description of the techniques used to gather the files for each file type is provided together with the actions that were performed on the files to confirm that they were of the highest quality and provided an accurate representation of their specific file type. Details are also provided on the content of the entire data set as well as instructions on how researchers can gain free and unlimited access to the final data set.

While the data set was initially created to aid research in ransomware detection, it is sufficiently broad and diverse enough to allow for its application in many other areas of research that require a varied mixture of common real-world file examples. The \emph{NapierOne} data set is an ongoing project and researchers are strongly encouraged to leverage this data set in their own research. 

\end{abstract}

\begin{keywords}
\sep Corpus \sep Mixed File Dataset \sep Govdocs1 \sep Malware \sep Ransomware \sep Entropy \sep Forensics
\sep
\sep
\sep
----------------------------------------------
\sep
Article history
\sep
Received 27\textsuperscript{th} May 2021
\sep
Accepted ?
\sep
Available
\end{keywords}

\maketitle
\section{Introduction}
\label{cha:intro}
\noindent To facilitate research repeatability, it is necessary that well respected, realistic and easily accessible standard data sets are used. This approach is confirmed by \cite{Garfinkel2010,Roussev2009} who stress that test data must be representative of data likely to be encountered in real-world situations. It is a known issue within the malware research community \cite{Al-rimy2018b,Grajeda2017,Maigida2019,Pont2019}, that there is often a lack of readily available, researchable ransomware data sets with Grajeda et al.~\cite{Grajeda2017} finding that only 4\% of malware researchers end up publishing their data sets.

The main contributions of this paper are firstly, the methodology for creating a mixed file data set that contains real-world examples of commonly used file formats, and secondly, the \emph{NapierOne} mixed file data set itself. The data set was developed during ransomware analysis research and with a focus on the development and testing of ransomware detection systems. A common characteristic of ransomware programs is that at some point they will attempt to encrypt the user's files. Tests can be used to determine the randomness of a file, with higher randomness, or entropy, suggesting that the contents may be encrypted. One drawback with this approach is, though, that the content of some legitimate files such as archive and image files can also appear to be random. Generally, the problem of detecting ransomware can be reduced to the problem of detecting random data being written to the file system or stored in memory~\cite{Pont2020}. File types that normally have high entropy and appear to contain random data were included in this data set for this specific reason.

\subsection{Cybersecurity test data sets}
\noindent For cybersecurity, test data sets can cover a wide variety of data types including malware samples, email data sets, network traffic, memory images, disk and device images. This paper will focus on data sets that contain collections of files of various types ~\cite{Grajeda2017} and will use the term 'mixed file data set' to refer to this type of data set. Abt and Baier~\cite{Abt2016} classify this type of data set as one which contains real-world data, is publicly available and manually curated.

Previously, in an attempt to address the issue of a lack of data sets, Garfinkel et al.~\cite{Garfinkel2009} created a publicly available Govdocs1 corpus. One part of which contains nearly one million files collected using random searches of the US \emph{.gov} domain. The resulting files are stored in 1,000 directories, each containing 1,000 files. This corpus can be obtained from \cite{govdocs,Garfinkel}. 

While the Govdocs1 corpus is a large, varied, well-known and highly cited, mixed file data set containing more than 40 different file types, recently some researchers have found it necessary to augment the corpus with additional files. A summary of these enhancements and their justifications are:

\begin{itemize}
    \item The files in the data set are now more than 12 years old and it is unknown how accurately their type and content, reflect current usage.
    \item Some file types that have gained popularity since the creation of the data set are not well represented. Specifically, regarding the newer Microsoft Office document types: XLSX; DOCX; and PPTX~\cite{Nguyen2014}. 
    \item Closer examination of the data set reveals that a class of files known as archives are under-represented~\cite{Pont2020} within Govdocs1. Archive files are normally compressed and are often used to collect multiple files together into a single file for easier portability or to reduce storage space requirements. Archive files often store directory structures, use error detection, and sometimes use built-in encryption. There are multiple types of archives currently in use, with varying properties and characteristics, however, the only type present of any significant size in the original data set is gzip.
    \item Modern image file types used in web design are missing (e.g. WEBP)~\cite{Pont2020}.
    \item No examples of encrypted files or files with little structure and very high entropy are present~\cite{Penrose2013}.
    \item Files that have high entropy, such as encrypted files, tend to have little recognisable structure~\cite{Penrose2013}. No examples of such files currently appear in the Govdocs1 data set. The inclusion of these types of files would prove beneficial when testing ransomware detection systems.
    \item The file types contained in the Govdocs1 data set only reflect the files that were found on the .gov domain, and do not necessarily represent the general popularity of the chosen file types. 
\end{itemize}

\subsection{Paper contribution}
\noindent In this paper, we introduce a complementary data set for the Govdocs1 corpus, known as \emph{NapierOne}~\cite{Davies2021}, which may be used to address the points made above. The Govdocs1 data set and the techniques used to create it provided an excellent template on how to create and curate the \emph{NapierOne} data set. However, no actual data from the original Govdocs1 data set is present in the proposed new data set.  

An overview of the methodology used in creating the proposed data:
\begin{enumerate}
    \item Research was performed to identify popular file formats that could be candidates for inclusion into the mixed file data set. Usage statistics from more than 10 independent sources were gathered, aggregated and collated. This research produces a final list of file types that were included in the data set.
    \item 5,000 example files of each of the identified file types were then sourced.
    \item Some additional actions were then performed on some of the gathered files in order to generate additional data subsets, for example, the creation of collections of files into archives.
    \item Robust validation was then performed on all of the files in the data set, including virus scanning, submission to virustotal~\cite{VirusTotal2019}, duplication removal and file format verification.
    \item Each data subset was then documented and curated, with a copy finally being placed for public use on the www.napierone.com website ~\cite{Davies2021}.
    
\end{enumerate}

The presence of particular files types in the final data set was determined from research into the general popularity of the file type and not by its presence in a particular source.

While the data set was initially created to aid research into ransomware detection, it is felt that it is sufficiently broad and diverse enough to allow its use in many other areas of research such a file analysis, compression technique comparison, image compression comparison and Microsoft Office file format analysis to name just a few examples.

\section{Related Work}
\noindent Cybersecurity-related research data sets can be classified into various types depending on the data they contain. The following five classifications have been proposed ~\cite{Abt2016,Garfinkel2009}; Malware Samples, Disk Images, Memory Images, Network Data and File Data. The remainder of this paper will concentrate on file type data sets. This type of data set contains a rich and varied collection of typical file types that may be found on modern computer systems and can be used to generate target file collections that are attractive to ransomware attacks. A recurring theme in many of the papers surveyed is that there exists a lack of publicly available mixed file data sets available for malware research and this is a large obstacle faced by ransomware researchers~\cite{Al-rimy2018}.  Currently, many studies build their own data sets by downloading raw samples from public repositories. 
However, many of these studies do not follow standard approaches for creating these data sets. Building a high quality, publicly available data set could be of great use, as it would contribute to building robust and accurate detection models.

Berrueta~\cite{Berrueta2019} identifies that there are no common metrics of accuracy and performance in ransomware detection and the fragmentation of scientific research on ransomware combined with a lack of coherent investigation methodology is a major challenge in this research~\cite{Dargahi2019}. This view is supported by Maigida~\cite{Maigida2019} who state that the lack of readily available, researchable ransomware data sets is also hindering the speedy development of detection and prevention solutions. 
The availability of up-to-date ransomware data sets is critical in evaluating newly proposed detection methods as the advancement of ransomware techniques could quickly render old data sets irrelevant. Almost no reviewed proposal for a ransomware detection technique provided enough detail on how the data set was created or sufficient description of its actual content to allow it to be recreated by other researchers interested in reconstructing their environment and validating the research results.

It has been reported by Grajeda et al.~\cite{Grajeda2017} that as many as two-thirds of the data sets used by researchers are experimentally generated with only a third being created from real-world data. They go on to say that in 96\% of cases, these data sets are not released for public use or scrutiny. This is against the recommendations of some researchers in the field~\cite{Abt2016,Fitzgerald2012,Grajeda2017} who stress the importance of sharing data sets, allowing researchers to replicate results and improve the state-of-the-art~\cite{Grajeda2017}. Building a public, ready-to-use ransomware data set would facilitate upcoming studies~\cite{Grajeda2017} and encourage more researchers to further investigate ransomware and produce solutions for various issues~\cite{Grajeda2017}. Two separate independent surveys agree with this conclusion, finding that these data sets would contribute to building robust and accurate detection models~\cite{Al-rimy2018,Pont2019}. In addition to this, the development of a universal testing platform allowing each of the methods to be evaluated in isolation using the same data sets would provide transparency in the evaluation processes as each test would be performed in a consistent, reproducible manner allowing direct comparisons to be made~\cite{Garfinkel2009,Penrose2013,Pont2019}. Berrueta~\cite{Berrueta2019} goes further and states that comparability and reproducibility are neither facilitated by the problem at hand nor by the way researchers present their results.

The research performed by Grajada~\cite{Grajeda2017} shows that many researchers prefer not to share their data sets~\cite{Abt2016}. Offered reasons for which being, firstly, researchers may not have the capability of sharing the set due to its size and the researcher's lack of available resources. A second factor may be related to the data set content as well as law and privacy concerns. Thirdly, the importance of sharing their data had not been considered. Finally, manually compiling data sets can be time-consuming, sometimes requiring months of work, so researchers having access to data sets with limited public accessibility have a clear competitive advantage.  Grajeda et al.~\cite{Grajeda2017} research goes on to show that less than 4\% of researchers shared their data set while on the other hand almost 50\% make use of existing data sets. In other words, whenever a repository or a sophisticated data set is available, researchers appreciate and utilise it. 

The following defines some of the publicly available mixed data sets. 

\subsection{Govdocs1}
\noindent The most well-known publicly available data set used in malware analysis today was developed by Garfinkel~\cite{Garfinkel2009,Garfinkel} in 2009 and is known as Govdocs1. The data set was designed to enable reproducibility of forensic research but makes no claims regarding the popularity of the file types it contains.
One part of the data set contains approximately one million files collected using random searches of the \emph{.gov} domain. It has been used by many ransomware researchers~\cite{Cleary2018,Fitzgerald2012,Grajeda2017,Kolodenker2017,McCarrin2013,Nguyen2014,Penrose2013,Pont2019,Pont2020,Roussev2011,Roussev2013}, supporting the claim that this data set is a well-known and respected source of test data.  In 2017 Grajeda et al.~\cite{Grajeda2017} reported that this was the most popular data set currently in use. The files are stored in 1,000 directories, each containing 1,000 files of various types, as well as 10 randomly assigned streams for development and testing purposes. 

This data set is now more than ten years old and during this time the use of some file types has diminished and new types have emerged. The sample size of some file types within the corpus now does not accurately reflect modern use. For example, there exist only 169 examples of files with the DOCX format, whereas, in reality, this format has become prevalent. Microsoft Office document file types, such as XLSX, DOCX and PPTX, are often targeted by ransomware~\cite{Al-rimy2018,Jung2018,Kharraz2016}, so their inclusion in the data set would be beneficial in ransomware research testing. A table illustrating the file types present in this corpus together with its sample size is shown in Table~\ref{tab:govdocstypes}.

\begin{small}
\begin{table}
\setlength{\tabcolsep}{4pt}
\caption{Govdocs1 File Types and File Counts}
\centering
\begin{tabular}{llllllll}
\toprule
\textbf{Ext} & \textbf{\# files} & \textbf{} & \textbf{Ext} & \textbf{\# files} & \textbf{} & \textbf{Ext} & \textbf{\# files} \\
\midrule
bmp   & 75 &      & jar & 34 &       & sql & 632    \\
chp   & 2 &       & java & 323 &     & squeak & 1   \\
csv   & 18,396 &  & jpg & 109,281 &  & swf & 3,691  \\
data  & 1 &       & js & 92 &        & sys & 8      \\
dll   & 7 &       & kml & 995 &      & tif & 3      \\
doc   & 80,648 &  & kmz & 949 &      & tmp & 196    \\
docx  & 169 &     & log & 10,241 &   & ttf & 104    \\
dwf   & 474 &     & pdf & 232,791 &  & txt & 84,091 \\
eps   & 5,465 &   & png & 4,125 &    & wp & 17      \\
exe   & 5 &       & pps & 1,629 &    & xbm & 51     \\
exported & 5 &    & ppt & 50,257 &   & xls & 66,599 \\
gif   & 36,301 &  & pptx & 219 &     & xlsx & 46    \\
gz    & 13,870 &  & ps & 22,129 &    & xml & 41,994 \\
hlp   & 660 &     & pst & 1 &        & zip & 27     \\
html  & 191,409 & & pub & 76 &  &  & \\
\bottomrule
\end{tabular}
\label{tab:govdocstypes}
\end{table}
\end{small}

Apart from the small data set size of modern Microsoft Office formats, there are also few examples of different archive types and other highly entropic file types and no examples of encrypted files. Entropy is often used in ransomware detection systems, so the presence of high entropy files in the data set, would be useful during the development and testing of such systems. 
To address these points, some researchers~\cite{Nguyen2014,Penrose2013,Pont2020} have attempted to enrich the original data set with additional file types, before using it, making the resulting hybrid data set more realistic and relevant to their research.

\subsection {t5}
\noindent When evaluating ssdeep and sdhash for similarity matching, Roussev~\cite{t5corpus,Roussev2009} created the \emph{t5} corpus based on the first four directories of the Govdocs1 corpus. Files that were smaller than 4KB and larger than 16.5MB being excluded from this data set. The resulting corpus has 4,457 files including DOC, GIF, HTML, JPG, PDF, PPT, TXT and XLS. Approximately 45.7\% of the files are text-based (TXT or HTML). The files were selected from neighbouring directories in the Govdocs1 corpus in the hope that there would be some similarity between them as they are likely to have originated on the same server. As this data set is based on the Govdocs1 data set, the same issues also exist with this data set such as that there are only a limited number of modern file types present. The t5 corpus has been used extensively in research into approximate matching~\cite{Chang2015,McCarrin2013,Roussev2011,Scaife2016}, which is a technique used to identify the similarity between digital artefacts (sequence of bytes)~\cite{Cleary2018}.

\subsection {msx-13}
\noindent Continuing in his research, Roussev~\cite{Roussev2013} developed the \emph{msx-13} corpus containing over 22,000 Microsoft Office 2007 documents. DOCX, XLSX and PPTX files were gathered using Google queries across 10 domains which favour English language content: .com, .net, .org, .edu, .gov, .us, .uk, .ca, .au and nz. The data set also contains MSZ archive files which are zip containers containing deflate-encoded content and embedded objects in their native format, however, they found great variations in the content of the files making it impossible to come up with a general classification scheme. It was also noted that the gathered PPTX files were much larger and had many more embedded objects than DOCX and XLSX.

\subsection {Other Sources}
\noindent Some researchers describe in detail how they generated their data sets. For example, Pont~\cite{Pont2020} outlines the precise technique they used to generate their data. Initially, their data set was based on the Govdocs1 corpus but then enriched with multiple image types and archive files~\cite{pontdataset}. This appears to be the most up-to-date comprehensive data set available, but due to its recent release, it is unknown if it has been leveraged by other researchers. Jung~\cite{Jung2018} also developed a data set that while it is only limited to PDF files, it does contain encrypted versions of these files as well. The methods used to generate the data set used by DeGaspari~\cite{DeGaspari2020} are also well described, discussing the files types present and their distribution, but the actual data set has not yet been made public by the researchers and the description is not sufficiently detailed enough to be able to reproduce the data set. During the research, the following websites were also searched for prospective data set examples~\cite{govdocs,UniversityofNewHaven2021}.

\subsection{Data Set Conclusions}
\noindent Considering all of this evidence from the reviewed literature, it is clear that data sets used within the ransomware testing community are currently facing several issues. These include the fact that many researchers are manually creating their own data sets which are often not released after the work is completed~\cite{Grajeda2017} and there is a lack of standardised data sets that are suitable for ransomware research~\cite{Abt2016,Al-rimy2018}. These weaknesses combined produce one of the major disadvantages facing the research community, that of low reproducibility and comparability~\cite{Grajeda2017}.

If a modern, diverse, representative mixed file data set could be created and made publicly available, its effect would be to improve the quality and pace of research especially in domains such as ransomware analysis and digital forensics~\cite{Grajeda2017}. Building such a data set would be of great use, as it would contribute to building robust and accurate detection models~\cite{Al-rimy2018}. Experimenting with real-world data is crucial for developing reliable algorithms and tools as ``how can we learn from our past when we do not have real, accessible data to learn from?'' \cite{Baggili2015}.

\section{Methodology}
\noindent Some of the forensic corpora reviewed in the previous section have proven to be an excellent resource in the field of ransomware research. The philosophy guiding the development of the \emph{NapierOne} data set was to mimic their success, by complementing these data sets, with one that contains more  examples of data formats that have become more prominent, as well as including files that have similar characteristics to encrypted files.
The techniques used in determining the data set candidate files types are discussed below followed by how the actual files for each type were sourced.

\label{cha:methodology}
\subsection{File Selection}
\label{subsection-file-selection}
\noindent An important aspect of building a representative data set relates to file type usage and popularity. It is known that Google gathers statistics on file types while it performs its website indexing searches. However, the statistics are only gathered on a limited number of file types~\cite{Google2015,Google2016}. 
While it was not possible to discover a definitive ranked list of files types currently in use, it was decided to adopt a consensus approach. This involved querying various sources of possible usage information and gathering approximate lists of up to their Top 40 file types. These lists were then compared and aggregated with the lists gathered from other sources, resulting in a fair representation of what are currently popular file types are in use today. The list produced is not proposed as definitive but rather a best guess consensus.
A list of the sources for file type usage information is discussed below. The alphabetical identification used for each data source is then repeated in the data set description shown in Table~\ref{tab:data-set-details}.

\noindent\textbf{A - VirusTotal} Statistics taken from file submissions to this website~\cite{VirusTotal2019} during that past 12 months were recorded.\\
\noindent\textbf{B - W3Techs} This website~\cite{W3Techs2014} provides information about the usage of various types of technologies on the web. Statistics relating to the most popular images types were gathered from here.\\
\noindent\textbf{C - Search Engine Statistics} Searches were performed for more than 100 different candidate file types. Search trends and statistics were also analysed and recorded in an attempt to identify the popularity of the file type. Although Google only index certain file types~\cite{Google2015}, their justification for doing so could also be considered a reason to include these files types in the data set.\\
\noindent\textbf{D - File Preservation} There exists research into the investigation of digital data preservation. One aspect of this research is to identify popular file formats. The findings from this research~\cite{dpconline,libraryofcongress,smithsonian,Jackson2012,Ryan2014} were also taken into account, as this research identifies certain files as being important and have the potential for remaining in active use for the foreseeable future.\\
\noindent\textbf{E - MIME Type Statistics} A MIME type~\cite{Freed1996,Freed2013} is a standard label used to indicate the nature and format of a document, file, or assortment of bytes. It is mainly used in Internet-based technology so that software can know how to handle the data. It serves the same purpose on the Internet that file extensions do on Microsoft Windows. When crawling the internet, these labels may be recorded and the frequency of them can be used to identify the popularity of file formats. Statistics have thus been gained from the WayBackMachine~\cite{wayback} and for CommonCrawl~\cite{CommonCrawl2021}, who record and publish monthly statistics on MIME types that are encountered.\\
\noindent\textbf{F - fileinfo.com} This website~\cite{fileinfo} performs research into file types and regularly publishes ranked lists of popular file types. These popularity lists are calculated from both the requests that the website receives and general web traffic analysis.\\
\noindent\textbf{G - Ransomware Targets} One of the anticipated initial applications of this data set will be in the field of ransomware research. For the data set to be useful in this research, it should contain a good selection of file types that are regularly targeted by ransomware~\cite{Al-rimy2018,Jung2018,Kharraz2016}, typical examples of which being DOCX and XLSX. This additional criterion did not impact the main goal of listing common file types as these file types appear in multiple lists.\\
\noindent\textbf{H - High Entropy} As the data set will initially be used in ransomware detection testing, it was considered a benefit for the data set to contain files that had this attribute. One technique used to identify files that have been affected by ransomware is to test the file's entropy so including files that normally have this attribute, such as archive files, would provide realistic challenges in ransomware identification. This secondary consideration does not impact the overall goal of generating the file type popularity list.\\
\noindent\textbf{I - File-extension} These are others websites~\cite{fileextension,File-extensions.org2021} specialising in file type research and regularly publish file type usage statistics.\\
\noindent\textbf{J - OS Installation} File type count statistics were gathered from clean windows installations (XP 2003 SP1, Vista 2006 and Windows 10 20H2) as well as four clean 64Bit Linux installations (MX Linux 19.4, Debian 5.9.1, Ubuntu 18.04.1, and Ubuntu 20.10).\\
\noindent\textbf{K - Govdocs1} Inclusion of the file types in the original Govdocs1 data set~\cite{govdocs} were also taken into consideration when deciding if they should appear in this data set. However, the actual files being sourced from elsewhere.\\

The results for all the sources were then sorted by popularity and aggregated into a list of approximately 40 file types. To encourage consensus, the file types needed to appear in at least two lists before they were considered in the final results.

The contents of the new data set were generated in the following ways:
\begin{itemize}
    \item Examples of files using the most commonly used document types such as DOC, DOCX, PDF, PPT, PPTX, XLS, and XLS  were be gathered from the UK government domain, \emph{gov.uk}. Files collected from this domain are covered by the 'Open Government License agreement for Public Sector Information'~\cite{OGL}. This agreement allows for the free distribution of the documents as long as they are accompanied by the correct attribution.
    \item Image files in the TIF format were taken from the RAISE data set~\cite{Dang-Nguyen2015}. This data set permits redistribution of its contents as long as it is accompanied by the correct attribution. Subsequent image file type data subsets, for example, BMP, DWG, EPS, GIF, JPG, PNG, SVG and WEBP were generated from this original TIF data set. In addition, examples of image files for each of these image types were taken from the CommonCrawl~\cite{CommonCrawl2021} dataset, thus providing real-world examples of these file types.
    \item Archive data subsets were generated using files from the documents subset.
    \item Audio files were gathered from the Free Music Archive ~\cite{Defferrard2017}. Its license permits free and unrestricted usage and distribution.
    \item Video files, in MP4, format, were gathered from the Kinetics-700 data set~\cite{Carreira2020,Deepmind2017,Gitumarkk2019}. Its license also allows free usage and distribution when using the correct attribution. Other video data sets in different formats were generated by transcoding this MP4 data set. Real-world examples of video files were also sourced from the CommonCrawl~\cite{CommonCrawl2021} dataset..
    \item Examples of files in the EPUB format were taken from Project Gutenberg~\cite{Hart2021}, which is an online library of free eBooks.
\end{itemize}

\subsection{Ethics}
\noindent Current research institutional guidelines concerning the handling of Personally Identifiable Information (PII) present within data sets are,  that if the data has been publicly sourced, as the files in the data set have been, there is no requirement for further anonymising or obfuscation of the data. The justification for this being that the information is already in the public domain. This is considered to be a positive advantage as this data set would then be useful for researchers that rely on the file metadata being present. As well as also preserving the data in its original form.

However, there remains a possibility in the future, depending on changing guidelines or regulations which may require the data to be cleaned. Possible future actions that could be performed on the data could be to have any versioning, comments and metadata removed. An attempt to anonymise any PII from the content of the files could also be undertaken if required. To date, these actions have not yet been performed on the data set.
 
In accordance with organisation guidelines and best practice recommendations \cite{Grajeda2017, Sharpless2015} a Data Management Plan (DMP) has been developed for this data set and is available on request. This report contains precise information relating to the data set's management, curation, retention, security, access and responsibilities. Attribution information is provided with each data subset and these should be followed when the subset is used.
\section{Data Set Creation}
\label{cha:data-set-creation}
\noindent Over 40 distinct file types were identified from the research described in Section~\ref{subsection-file-selection}. Some file type data sets were further divided into multiple data subsets. These subsets were used to contain examples of files that demonstrate different characteristics. For example, multiple data subsets were created for the ZIP file type. Each subset containing examples of files compressed using different compression algorithms. In most case, there exist 5,000 example files for each data set and subset.
The resulting \emph{NapierOne}~\cite{Davies2021} data set contains nearly 500,000 unique files distributed between 100 separate data sets and subsets. A detailed overview of the structure of the data set is shown in Figure~\ref{fig:data-set-sources}.

\subsection{Data Set Structure}
\begin{figure}
  \includegraphics[width=0.9\columnwidth]{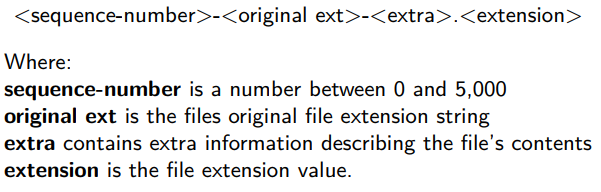}
  \caption{File naming convention}
  \label{fig:filename-format}
\end{figure}

\noindent As with the Govdocs1 data set, once the data had been downloaded it needed to be analysed, validated, characterised and curated~\cite{Garfinkel2009}. To provide consistency across the entire data set, a standardised naming convention was used. Reasons for renaming the files to follow this convention were: firstly the file names reflect to some extent the content and structure of the file itself and so are self-documenting. The renaming processes provides a small degree of anonymisation and finally, files containing related content are linked via their sequence numbers. A description of the file naming convention used is shown in Figure~\ref{fig:filename-format}. The inclusion of the sequence number facilitates the possibility of being able to cross-reference certain related files across data subsets, such as in tracing image files or archives that use the same underlying content. The file's extension value appears as part of the file name, as well as the file's actual extension. It is in both places because in some cases, such as when a file is encrypted by ransomware, the actual extension is modified. In these circumstances, the file's original extension can still be seen from the file name. 

\begin{figure}
  \includegraphics[width=0.9\columnwidth]{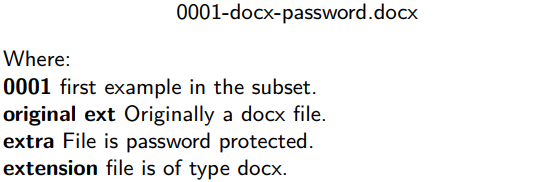}
  \caption{Example file name.}
  \label{fig:filename-example}
\end{figure}

For example, for the first file in the DOCX data subset that has been password-protected, the name would be the value shown in Figure~\ref{fig:filename-example}. Examples of how the sequence number is used to link related files are shown in Tables~\ref{tab:image-naming-relationship} and ~\ref{tab:similar-naming-archive}. The authors retain the ability to determine each file's original file name if required.
When renaming the files, a record is kept of the mapping between the original and new file names to allow the original name to be determined if required.

The files within the data set are organised by file type into a directory hierarchy. On the top level are the main file types, such as DOC, PPTX, ZIP, and so on. Below these main type classification directories are one or more subdirectories, one for each type of file format variation. For example, a BZIP directory containing files created using the bzip compression method of the ZIP program, would be found under the main zip directory. The directory structure of the data set can be deduced from Table~\ref{tab:data-set-details}. Contained within each directory that holds data, is a description file in both HTML and PDF formats, which describes the data held within that data subset. The description file includes amongst other things: a description of the data files; where the data was sourced; the licence details; and contact information.

\subsection{File Sources}
\noindent The \emph{NapierOne}~\cite{Davies2021} data set is a collection of multiple data subsets each associated with one specific file type or a specific, configuration of a file type. However, on a conceptual level, the individual data subsets can be grouped as follows.\\

\textbf{Document files}. Various techniques were employed to gather file examples from websites covered by the \emph{gov.uk} domain. These included using the API functionality provided by the following gateways~\cite{Calderdale,Ckan,Data.gov.uk}, direct data set download and files available for download on sub-domain websites. Using these techniques it was possible to gather examples of files with the following file formats: CSV, DOC, DOCX, JPG, PDF, PPT, PPTX, TXT, XLS, XLSX and WEBP. Two additional data subsets were also created for each of these main file types. One where the magic number for the file has been removed and a second where the original file is protected with a password. The same password being used for all files, the password used was \emph{napierone}~\cite{Davies2021}. Data sets for the OpenOffice file formats, ODS and OXPS, were derived from the Microsoft Office files. Some other web-related files such as CSS, HTML, Javascript, JSON and XML were also sourced from these websites. ICS calendar data was generated using the following program~\cite{AntonioFranco}.\\

\begin{figure}
  \includegraphics[width=\columnwidth]{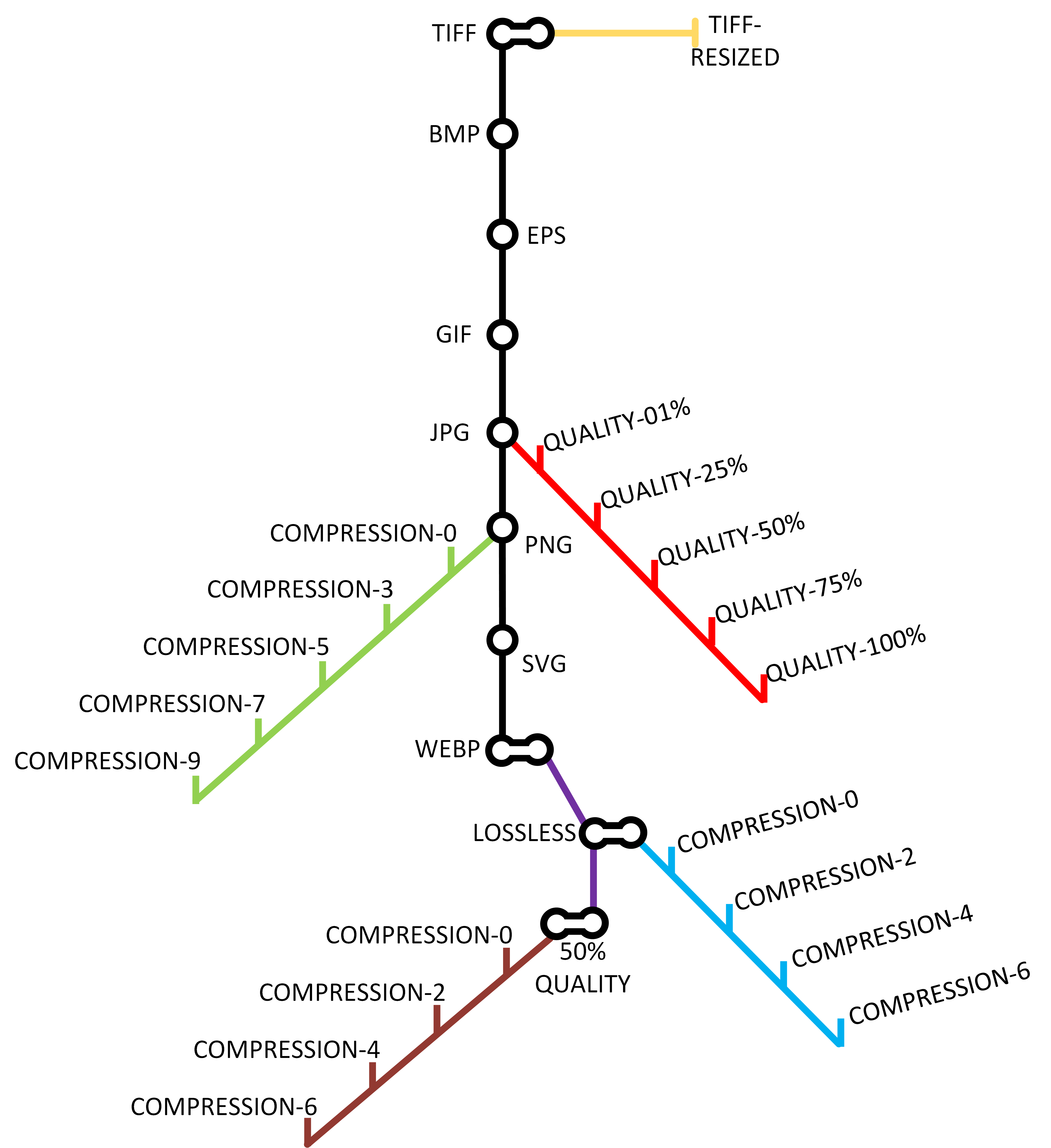}
  \caption{Image File Relationships.}
  \label{fig:image-data-subset-relationship}
\end{figure}

\textbf{Media files}. Two distinct methods were used to gather files of these types. Firstly real-world examples were retrieved from the the CommonCrawl~\cite{CommonCrawl2021} dataset. Secondly, random image files in an uncompressed format were gathered from the RAISE~\cite{Dang-Nguyen2015} public archive to create an initial image data subset. A second data subset was then generated in the same format, based on the original data set but with a reduced size. Images in the original TIF data subset were then converted into different image formats using the Advanced Batch Image Converter program~\cite{Hiestand2021}. Image data subsets derived from the original TIF image subset were BMP, DWG, EPS, GIF, JPG, PNG, SVG and WEBP. Some image subsets have multiple examples of a particular file type, for example, JPG and WEBP, with differing encoding qualities and compression levels. A graphical representation of this relationship is shown in Figure~\ref{fig:image-data-subset-relationship}. One side effect of this file relationship being that image files that have the same sequence number will contain a rendering of the same image. Using the naming convention outlined in Figure~\ref{fig:filename-format}, the relationship that the original TIF image \emph{0001-tif.TIF} has with all the other image formats is shown in Table~\ref{tab:image-naming-relationship}. If viewed, then all these files will display the same image. The JPG and WEBP data subsets were also augmented with images gathered from the gov.uk domains.

\begin{table}
\caption{Image File Relationship to Original TIF Image.}
\resizebox{\columnwidth}{!}{%
\begin{tabular}{l@{}|l}
\hline
 \textbf{Filename} & \textbf{Description} \\
\hline
0001-tif.TIF         & Original image taken from RAISE in TIF format\\
0001-tif-resized.TIF & Reduced size version in TIF format\\
0001-webp-lossless-c0.webp & Lossless compression level 0 version in WEBP format\\
0001-webp-lossless-c2.webp & Lossless compression level 2 version in WEBP format\\
0001-webp-lossless-c4.webp & Lossless compression level 4 version in WEBP format\\
0001-webp-lossless-c6.webp & Lossless compression level 6 version in WEBP format\\
0001-webp-q50-c0.webp & 50\% quality compression level 0 version in WEBP format\\
0001-webp-q50-c2.webp & 50\% quality compression level 2 version in WEBP format\\
0001-webp-q50-c4.webp & 50\% quality compression level 4 version in WEBP format\\
0001-webp-q50-c6.webp & 50\% quality compression level 6 version in WEBP format\\
0001-svg.svg & Version of original image in SVG format\\
0001-png-c0.png & Compression level 0 version in PNG format\\
0001-png-c3.png & Compression level 3 version in PNG format\\
0001-png-c5.png & Compression level 5 version in PNG format\\
0001-png-c7.png & Compression level 7 version in PNG format\\
0001-png-c9.png & Compression level 9 version in PNG format\\
0001-jpg-q001.jpg & 1\% quality compression version in JPG format\\
0001-jpg-q025.jpg & 25\% quality compression version in JPG format\\
0001-jpg-q050.jpg & 50\% quality compression version in JPG format\\
0001-jpg-q075.jpg & 75\% quality compression version in JPG format\\
0001-jpg-q100.jpg & 100\% quality compression version in JPG format\\
0001-gif.gif & Version of original image in GIF format\\
0001-dwg.dwg & Version of original image in DWG format\\
0001-eps.eps & Version of original image in EPS format\\
0001-bmp.bmp & Version of original image in BMP format\\
\hline
\end{tabular}
}
\label{tab:image-naming-relationship}
\end{table}

Examples of other media data subsets present are MP3 audio files taken from the Free Music Archive~\cite{Defferrard2017} and MP4 video files taken from the Kinetics-700 archive~\cite{Carreira2020,Deepmind2017,Gitumarkk2019} as well as sourcing real-world examples from the CommonCrawl~\cite{CommonCrawl2021} dataset. Examples of files in the e-book format EPUB were gathered from the project Gutenburg~\cite{Hart2021} which is a library of over 60,000 free eBooks. \\

\textbf{System files}. Data subsets containing files relating to system installation are also present. Files relating to Linux installations (ELF, BIN) were gathered from clean Linux installations of Debian 5.9.1, MX Linux 19.4, Ubuntu 18.04.1 and Ubuntu 20.10. DLL and EXE file examples were taken from clean Windows XP and  Vista installations. A similar action was performed, on all the program files gathered from operating system distributions, as Garfinkel~\cite{Garfinkel2009,Garfinkel2009a} applied to their executable files. That was to alter a portion of a file so that the original file could not be identified and could not be executed. The alteration was made in such a way as to minimise the effect on the entropy of the file. This alteration allows analysis of executable files but prevents them from being used without a license, which is believed to be sufficient redaction for the
purpose of distributing the files under the \textit{'fair use'} provisions of the US Copyright Act.
Some other configuration file types such as CSS and JSON were also sourced in this way.
Android installation files, APK, were gathered from the AndroZoo public data set~\cite{UniversiteduLuxembourg2021}. \\

\textbf{Other files}. Other data subsets present are a pseudo-random data subset named PSEUDO. This is a synthetic data subset of files created using the Python pseudo-random number generator function \texttt{'os.urandom'}. Files were created using this technique with lengths varying between 512 and 2,048 bytes. A second random number data subset called PURE is present which has files that contain true random numbers generated using unpredictable physical means, for example, atmospheric noise~\cite{random}. Files were created using this technique with lengths also varying between 512 and 2,048 bytes. Data sets of these types prove useful as a benchmark when comparing against other highly entropic files such as archives.

File type with the DAT extension describe files containing some sort of text or binary data. Formatting of this data depends on the program which created or utilises the given file and can vary in content and structure. As it was not possible to gather a representative example of this file type, it was not included in the final data set. The TAR file format was not that popular, but its presence was required as an intermediate data subset when creating other data subsets such as ZLIB. It is included in this data set for this reason.

\subsection{File Structure}
\noindent Once the main data types had been collected, two additional actions were performed. Firstly a directory structure was generated where 5,000 sub-directories were created named from 0001 to 5000. Each of these sub-directories was then populated with examples from the document and image data subsets, resulting in each directory containing between 5 and 15 files of various types. The purpose of this file structure was to form a hierarchy that could be used when creating data subsets of archive file types. The following archive data subsets were then created by performing compression on this file structure: 7Zip, GZIP, RAR, TAR, Zip and ZLib. As these file formats tend to have high entropy, they will be useful in ransomware detection testing. For some file types, multiple data subsets were created allowing the option to represent different compression levels or compression techniques. Again the same naming convention was leveraged, so using the sequence number, individual-related archives can be linked together. For example, Table~\ref{tab:similar-naming-archive} shows the different archives that were generated from the source directory 0001. Unless otherwise stated, the compression tool's default values were used in the creation of the archives. 

\begin{table}
\caption{Archive File Relationships}
\resizebox{\columnwidth}{!}{%
\begin{tabular}{l@{}|l}
\hline
 \textbf{Filename} & \textbf{Description} \\
\hline
0001-7z-bzip2.7z           & 7zip compression using bzip method\\
0001-7z-encrypted.7z       & 7zip compression \& password protected\\
0001-7z-hightcompress.7z   & 7zip using high compression\\
0001-7z-lzma.7z            & 7zip compression using lzma method\\
0001-7z-lzma2.7z           & 7zip compression using lzma2 method\\
0001-7z-ppmd.7z            & 7zip compression using ppmd method\\
0001-gz.gz                 & gzip compression\\
0001-rar.rar               & RAR compression\\
0001-tar.tar               & TAR compression\\
0001-zip-bzip2.zip         & Zip compression using bzip method\\
0001-zip-deflate.zip       & Zip compression using DEFLATE \\
0001-zip-encrypted.zip     & Zip compression \& password protected\\
0001-zip-highcompress.zip  & Zip using high compression\\
0001-zip-lzma.zip          & Zip compression using lzma method\\
0001-zip-ppmd.zip          & Zip compression using ppmd method\\
0001-zlib.zlib             & Zlib compression\\
\hline
\end{tabular}}
\label{tab:similar-naming-archive}
\end{table}

\subsection{File Collection}
\noindent A single, file collection, directory was created which contained examples of file types that are typically targeted by ransomware \cite{Al-rimy2018,Jung2018,Kharraz2016}. Examples of these file types being DOC, DOCX, PDF, PPT, PPTX, XLS and XLSX. This directory would act as a ransomware target directory and was placed on a previously prepared, isolated test machine, and a ransomware sample was executed in an ethical manner. Valid ransomware samples being sourced from VirusTotal~\cite{VirusTotal2019}. Once the ransomware execution had completed, the target files were examined to determine if they had been affected by the ransomware's execution. If they had, then, the affected files were checked for viruses and after confirming that they were safe, were placed in their own data subset within the \emph{NapierOne} data set. Currently, data subsets of ransomware encrypted files have been created for the following ransomware strains: NotPetya; Sodinokibi; Maze; Phobos; Netwalker; Dharma; and Ryuk.

\begin{figure*}
\centering
  \includegraphics[width=0.75\textwidth]{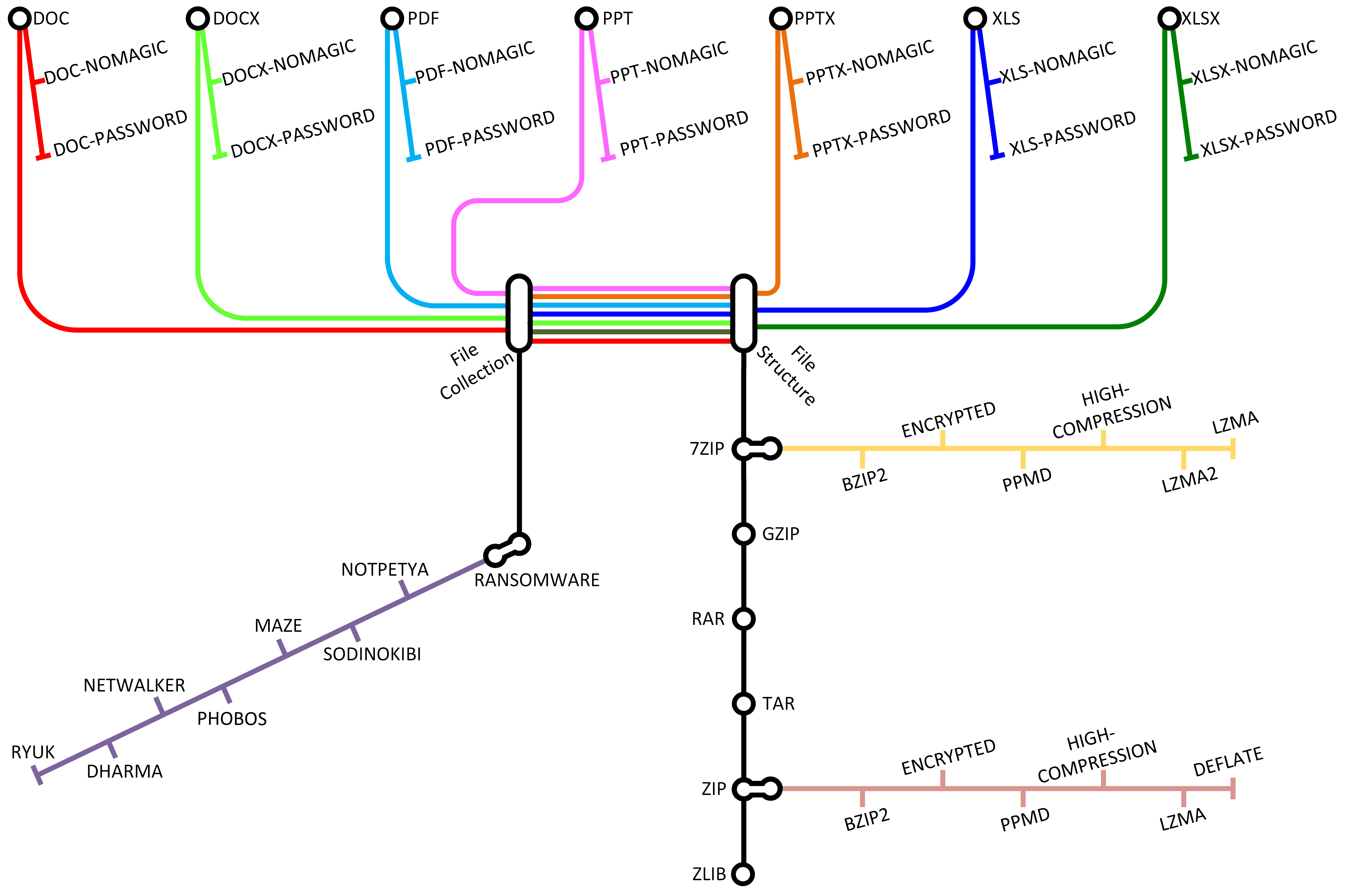}
  \caption{gov.uk File Collection Relationships.}
    \label{fig:gov-doc-file-used}
\end{figure*}

\begin{figure*}
  \includegraphics[width=\textwidth]{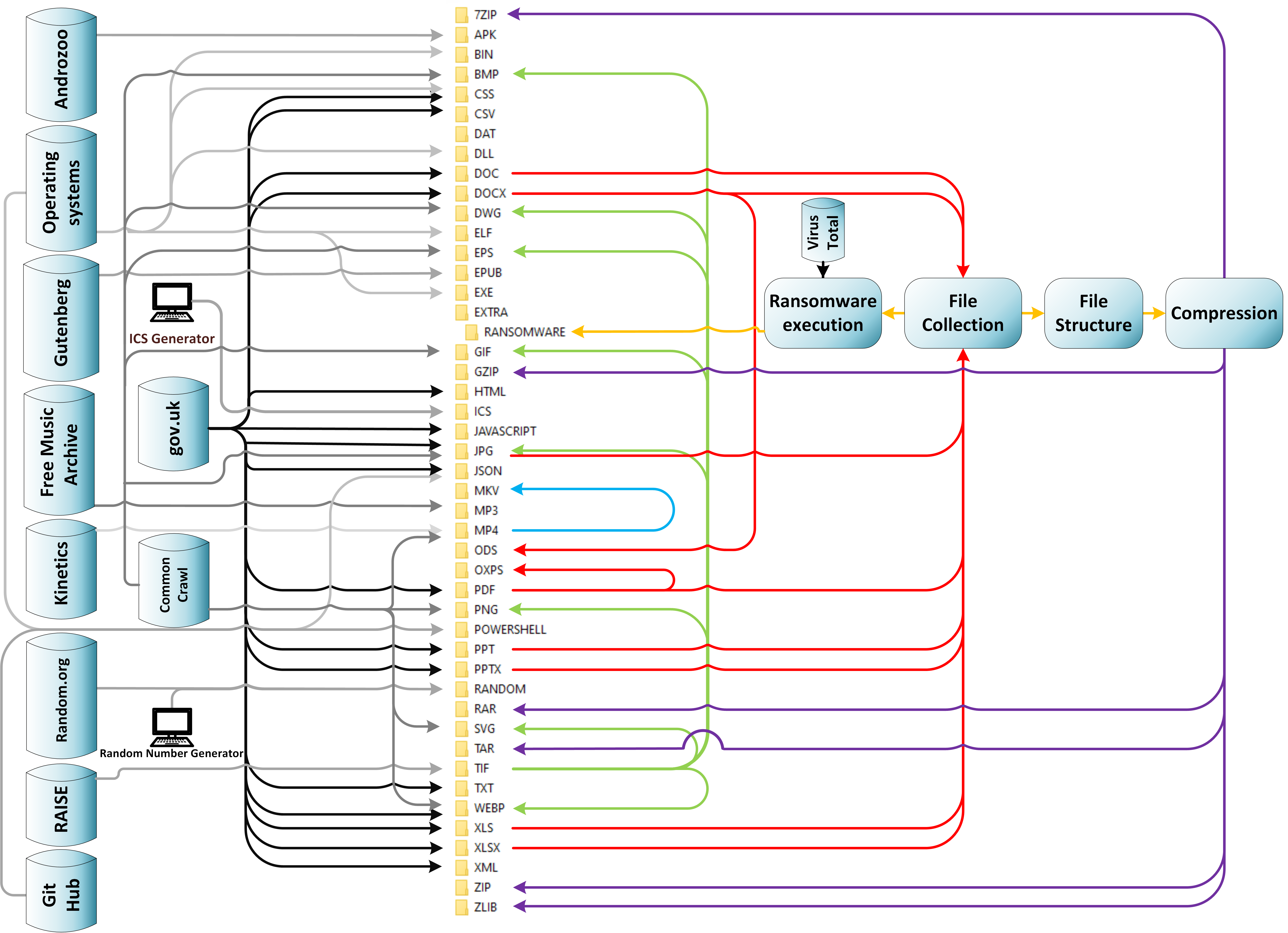}
  \caption{Data Set Sources.}
  \label{fig:data-set-sources}
\end{figure*}

Again leveraging the sequence number, the target file and ransomware encrypted file can be connected. A graphical representation of the relationships between the documents data subsets and the file collection and file structure hierarchies is shown in Figure~\ref{fig:gov-doc-file-used} and an overall view of the provenance of the individual data subsets is shown in Figure~\ref{fig:data-set-sources}. 

\subsection{File Validation}
\noindent Multiple actions were taken to ensure that the data contained within the data set was of the highest quality. File validation actions that were performed on all data files were:
\begin{itemize}
    \item All the files in the data set have been scanned for viruses using Windows Defender, details of which are given in Table~\ref{tab:programs-used}.
    \item The files were also checked to confirm that they contained data and that they have the correct magic numbers. 
    \item All the files were verified to confirm that they could be opened successfully using the appropriate application.
    \item Confirming that no duplicates existed in the data set. This was achieved by checking the files SHA1 hash value~\cite{Garfinkel2009}.
    \item Over 350,000 files that had been sourced externally as well as ransomware encrypted files were submitted to virustotal~\cite{VirusTotal2019} to confirm that they did not contain any malicious code.
    \item Image files were visually checked in an attempt to ensure that no illicit content was present.
\end{itemize} 

The tools used in the generation for the data set, its management and validation are described in Table~\ref{tab:programs-used}.
\begin{table}
\caption{Tools Used.}
\resizebox{\columnwidth}{!}{%
\begin{tabular}{l|l|l}
\hline
\textbf{Program names} & \textbf{Version} & \textbf{Comments}   \\
\hline
\makecell[l]{Advanced Batch Image \\Converter} & 1.2.2 &  \cite{Hiestand2021}   \\
Bulk Rename Utility & 3.3.2.0 &  www.bulkrenameutility.co.uk   \\

Antimalware Client & 4.18.2101.9 & Windows 10\\
Engine             & 1.1.17800.5 & Defender\\
Antivirus          & 1.331.1067.0 & \\
Anti-spyware       & 1.331.1067.0 & \\
PDF Reader         & 11.0.00 & Adobe\\
Microsoft Office   & 15.0.4430.2027 & Office Pro 2013\\
WinRAR             & 3.42 & www.rarlab.com\\
gzip               & 3.8.2 & Python module\\
7Zip               &  9.20 & www.7-zip.org\\
\hline
\end{tabular}}
\label{tab:programs-used}
\end{table}

\section{Evaluation}
\label{cha:evaluation}
\noindent A detailed description of the \emph{NapierOne} data set is presented in Table~\ref{tab:data-set-details}. The actual data set file system hierarchy levels can be deduced from the \emph{File Type}, \emph{Sub Classifications} and \emph{SubSub Classifications} columns. The data subset sizes are the size of the uncompressed data set. The justification for the file type being included in the data set is represented by the columns on the right-hand side of the table. The ranked lists where the specific file type appears is illustrated using the notation described in Section~\ref{subsection-file-selection}.

When creating the data subsets, the goal was for each collection to contain 5,000 example files. This was achieved in the majority of cases, with the only exception being file types in the Microsoft PowerPoint format (PPT, PPTX). The reduced size of this data set was due to the lack of example files that could be sourced from the \emph{gov.uk} domain. Instead of adding example files from other sources and mixing their provenance, it was decided to remain with the single source and accept that the data set population size was be reduced as a consequence.

\begin{table*}
\caption{NapierOne Mixed File Data Set}
\begingroup
\setlength{\tabcolsep}{6pt} 
\renewcommand{\arraystretch}{1.0685} 
    \tiny
    \centering
    \begin{tabular}{ l r l c c c c l l |p{0.05mm} p{0.5mm} p{0.5mm} p{0.5mm} p{0.5mm} p{0.5mm} p{0.5mm} p{0.5mm} p{0.5mm} p{0.5mm} p{0.5mm} }
        
        \toprule
        \multirow{1}{*}\textbf{File Type} & \multirow{1}{*}\textbf{Size (KB)} &  \multirow{1}{*}\textbf{Files \#}  &  \multirow{1}{*}\textbf{\makecell{Min\\Size(KB)}}  &  \multirow{1}{*}\textbf{\makecell{Max\\Size(KB)}}  &  \multirow{1}{*}\textbf{\makecell{Average\\Size(KB)}}  & \multirow{1}{*}\textbf{$\sigma$}  &  \multirow{2}{*}\textbf{Sub classifications} & \multirow{1}{*}\textbf{\makecell{SubSub\\ Classifications}} & \multicolumn{10}{c}{File Usage Source} \\
        & & & & & & & & & \textbf{A} & \textbf{B} & \textbf{C} & \textbf{D} & \textbf{E} & \textbf{F} & \textbf{G} & \textbf{H}& \textbf{I}& \textbf{J}& \textbf{K}\\ \toprule
        \arrayrulecolor{gray!40}
        APK &  9,310  &  5,000  &  14  &  3,000  &  2,001  &  778  & APK & - & $\blacksquare$  &  &  &  &  & $\blacksquare$  &  &  & $\blacksquare$  &  &  \\ \hline
        BIN &  3,346  &  5,000  &  10  &  164,411  &  669  &  5,512  & BIN & - &  &  &  &  &  & $\blacksquare$  &  &  & $\blacksquare$  & $\blacksquare$  &  \\ \hline
        BMP &  208,000  &  5,000  &  18,099  &  48,306  &  44,845  &  5,711  & BMP & - &  & $\blacksquare$  &  & $\blacksquare$  &  &  & $\blacksquare$  &  &  &  & $\blacksquare$  \\ 
         &  1,750  &  5,000  &  1  &  1,049  &  375  & 301   & BMP-FROM-WEB & - &  & $\blacksquare$  &  & $\blacksquare$  &  &  & $\blacksquare$  &  &  &  & $\blacksquare$  \\ \hline
        CSS &  355  &  5,000  &  1  &  4,698  &  74  &  234  & CSS & - &  &  &  & $\blacksquare$  & $\blacksquare$  &  &  &  &  & $\blacksquare$  &  \\ \hline
        CSV &  3,503  &  5,000  & 1 &  192,472  &  701  &  7,637  & CSV & - &  &  &  & $\blacksquare$  & $\blacksquare$  & $\blacksquare$  &  &  &  &  & $\blacksquare$  \\ \hline
        DLL &  3,130  &  5,000  & 1 &  155,445  &  673  &  4,219  & DLL & - & $\blacksquare$  &  &  &  &  &  &  &  & $\blacksquare$  & $\blacksquare$  & $\blacksquare$  \\ \hline
        DOC &  1,320  &  5,000  &  20  &  13,822  &  290  &  763  & DOC & - & $\blacksquare$  &  & $\blacksquare$  & $\blacksquare$  & $\blacksquare$  &  & $\blacksquare$  &  & $\blacksquare$  &  & $\blacksquare$  \\ 
         &  1,320  &  5,000  &  20  &  13,822  &  282  &  746  & DOC-NOMAGIC &  & $\blacksquare$  &  & $\blacksquare$  & $\blacksquare$  & $\blacksquare$  &  & $\blacksquare$  &  & $\blacksquare$  &  & $\blacksquare$  \\ 
         &  1,170  &  5,000  &  23  &  13,411  &  251  &  649  & DOC-PASSWORD & - & $\blacksquare$  &  & $\blacksquare$  & $\blacksquare$  & $\blacksquare$  &  & $\blacksquare$  &  & $\blacksquare$  &  & $\blacksquare$  \\ \hline
        DOCX &  1,420  &  5,000  &  11  &  14,580  &  303  &  866  & DOCX & - & $\blacksquare$  &  & $\blacksquare$  & $\blacksquare$  & $\blacksquare$  &  & $\blacksquare$  &  & $\blacksquare$  &  & $\blacksquare$  \\ 
         &  1,420  &  5,000  &  11  &  14,580  &  303  &  866  & DOCX-NOMAGIC & - & $\blacksquare$  &  & $\blacksquare$  & $\blacksquare$  & $\blacksquare$  &  & $\blacksquare$  &  & $\blacksquare$  &  & $\blacksquare$  \\ 
         &  1,590  &  5,000  &  18  &  14,707  &  311  &  872  & DOCX-PASSWORD & - & $\blacksquare$  &  & $\blacksquare$  & $\blacksquare$  & $\blacksquare$  &  & $\blacksquare$  &  & $\blacksquare$  &  & $\blacksquare$  \\ \hline
        DWG &  9,720  &  5,000  &  1,023  &  6,447  &  2,088  &  714  & DWG & - &  &  & $\blacksquare$  &  &  &  &  &  & $\blacksquare$  &  &  \\ 
         &  1,630 &  5,000  &  5  &  1,049  & 351 & 354 & DWG-FROM-WEB & - &  &  & $\blacksquare$  &  &  &  &  &  & $\blacksquare$  &  &  \\ \hline
        ELF &  1,560  &  5,000  &  1  &  130,513  &  312  &  3,064  & ELF & - & $\blacksquare$  &  &  &  &  &  &  &  & $\blacksquare$  & $\blacksquare$  &  \\ \hline
        EPS &  26,400  &  5,000  &  2,293  &  6,121  &  5,683  &  724  & EPS & - &  &  &  &  &  & $\blacksquare$  &  &  &  &  & $\blacksquare$  \\ 
         &  2,290  &  5,000  &  1  & 1,049  &  493  &  388  & EPS-FROM-WEB & - &  &  &  &  &  & $\blacksquare$  &  &  &  &  & $\blacksquare$  \\ \hline
        EPUB &  1,170  &  5,000  &  5  &  11,741  &  251  &  318  & EPUB & - &  &  &  & $\blacksquare$  & $\blacksquare$  & $\blacksquare$  &  &  & $\blacksquare$  &  &  \\ \hline
        EXE &  8,950  &  5,000  &  1  &  243,770  &  1,924  &  10,685  & EXE &  & $\blacksquare$  &  &  &  &  & $\blacksquare$  &  &  &  & $\blacksquare$  & $\blacksquare$  \\ \hline
        GIF &  3,030  &  5,000  &  108  &  1,172  &  650  &  141  & GIF & - &  & $\blacksquare$  &  & $\blacksquare$  & $\blacksquare$  &  & $\blacksquare$  &  &  & $\blacksquare$  & $\blacksquare$  \\ 
         &  913  &  5,000  & 1 & 1,049 & 192 &  312  & GIF-FROM-WEB & - &  & $\blacksquare$  &  & $\blacksquare$  & $\blacksquare$  &  & $\blacksquare$  &  &  & $\blacksquare$  & $\blacksquare$  \\ \hline
        GZIP &  18,200  &  5,000  &  6  &  59,749  &  3,909  &  4,858  & GZIP & - & $\blacksquare$  &  &  &  & $\blacksquare$  &  &  & $\blacksquare$  &  & $\blacksquare$  & $\blacksquare$  \\ \hline
        HTML &  303  &  5,000  &  1  &  6,804  &  64  &  174  & HTML & - & $\blacksquare$  &  & $\blacksquare$  & $\blacksquare$  & $\blacksquare$  &  & $\blacksquare$  &  &  & $\blacksquare$  & $\blacksquare$  \\ \hline
        ICS &  16  &  5,000  &  1  &  6,785  &  17  &  3  & ICS & - &  &  &  &  & $\blacksquare$  & $\blacksquare$  &  &  &  & $\blacksquare$  &  \\ \hline
        JAVASCRIPT &  86  &  5,000  &  1  &  6,785  &  17  &  130  & JAVASCRIPT & - & $\blacksquare$  &  &  & $\blacksquare$  & $\blacksquare$  &  & $\blacksquare$  &  &  & $\blacksquare$  & $\blacksquare$  \\ \hline
        JPG &  345  &  5,000  &  1  &  7,445  &  72  &  318  & JPG-FROM-WEB & - & $\blacksquare$  & $\blacksquare$  &  & $\blacksquare$  & $\blacksquare$  & $\blacksquare$  & $\blacksquare$  & $\blacksquare$  & $\blacksquare$  & $\blacksquare$  & $\blacksquare$  \\ 
         &  666  &  5,000  &  47  &  423  &  138  &  36  & QUALITY-01-PERCENT & - & $\blacksquare$  & $\blacksquare$  &  & $\blacksquare$  & $\blacksquare$  & $\blacksquare$  & $\blacksquare$  & $\blacksquare$  & $\blacksquare$  & $\blacksquare$  & $\blacksquare$  \\ 
         &  3,220  &  5,000  &  79  &  2,012  &  692  &  308  & QUALITY-25-PERCENT & - & $\blacksquare$  & $\blacksquare$  &  & $\blacksquare$  & $\blacksquare$  & $\blacksquare$  & $\blacksquare$  & $\blacksquare$  & $\blacksquare$  & $\blacksquare$  & $\blacksquare$  \\ 
         &  5,270  &  5,000  &  129  &  2,999  &  1,131  &  448  & QUALITY-50-PERCENT & - & $\blacksquare$  & $\blacksquare$  &  & $\blacksquare$  & $\blacksquare$  & $\blacksquare$  & $\blacksquare$  & $\blacksquare$  & $\blacksquare$  & $\blacksquare$  & $\blacksquare$  \\ 
         &  8,380  &  5,000  &  285  &  4,479  &  1,799  &  594  & QUALITY-75-PERCENT &  & $\blacksquare$  & $\blacksquare$  &  & $\blacksquare$  & $\blacksquare$  & $\blacksquare$  & $\blacksquare$  & $\blacksquare$  & $\blacksquare$  & $\blacksquare$  & $\blacksquare$  \\ 
         &  49,100  &  5,000  &  1,797  &  22,418  &  10,564  &  2,533  & QUALITY-100-PERCENT & - & $\blacksquare$  & $\blacksquare$  &  & $\blacksquare$  & $\blacksquare$  & $\blacksquare$  & $\blacksquare$  & $\blacksquare$  & $\blacksquare$  & $\blacksquare$  & $\blacksquare$  \\ \hline
        JSON &  202  &  5,000  &  1  &  70,217  &  137  &  2,439  & JSON &  &  &  &  &  & $\blacksquare$  & $\blacksquare$  &  &  & $\blacksquare$  & $\blacksquare$  &  \\ \hline
        MKV &  7,530  &  5,000  &  81  &  6,830  &  1,616  &  906  & MKV & - &  &  &  &  &  & $\blacksquare$  &  &  & $\blacksquare$  &  &  \\ \hline
        MP3 &  4,650  &  5,000  &  415  &  1,206  &  995  &  230  & MP3 & - &  &  &  & $\blacksquare$  & $\blacksquare$  & $\blacksquare$  & $\blacksquare$  &  & $\blacksquare$  &  &  \\ \hline
        MP4 &  7,540  &  5,000  &  81  &  6,933  &  1,619  &  906  & MP4 & - &  &  &  & $\blacksquare$  & $\blacksquare$  & $\blacksquare$  & $\blacksquare$  &  & $\blacksquare$  &  &  \\ 
         &  7,110  &  5,000  &  16  & 61,335   & 1,528   &  2,250  & MP4-FROM-WEB & - &  &  &  & $\blacksquare$  & $\blacksquare$  & $\blacksquare$  & $\blacksquare$  &  & $\blacksquare$  &  &  \\ \hline
        ODS &  1,500  &  5,000  &  9  &  48,620  &  322  &  1,513  & ODS & - & $\blacksquare$  &  & $\blacksquare$  &  & $\blacksquare$  &  &  &  & $\blacksquare$  &  &  \\ \hline
        OXPS &  1,320  &  5,000  &  20  &  13,822  &  282  &  746  & OXPS &  & $\blacksquare$  &  &  &  & $\blacksquare$  &  &  &  & $\blacksquare$  &  &  \\ \hline
        PDF &  4,510  &  5,000  &  3  &  56,986  &  968  &  2,799  & PDF & - & $\blacksquare$  &  & $\blacksquare$  & $\blacksquare$  & $\blacksquare$  & $\blacksquare$  & $\blacksquare$  &  & $\blacksquare$  &  & $\blacksquare$  \\ 
         &  4,510  &  5,000  &  3  &  56,986  &  968  &  2,799  & PDF-NOMAGIC & - & $\blacksquare$  &  & $\blacksquare$  & $\blacksquare$  & $\blacksquare$  & $\blacksquare$  & $\blacksquare$  &  & $\blacksquare$  &  & $\blacksquare$  \\ 
         &  4,220  &  5,000  &  2  &  56,903  &  906  &  2,724  & PDF-PASSWORD & - & $\blacksquare$  &  & $\blacksquare$  & $\blacksquare$  & $\blacksquare$  & $\blacksquare$  & $\blacksquare$  &  & $\blacksquare$  &  & $\blacksquare$  \\ \hline
        PNG &  175,000  &  5,000  &  10,941  &  47,713  &  37,792  &  5,664  & LOSSLESS & COMPRESSION-0 & $\blacksquare$  & $\blacksquare$  &  & $\blacksquare$  & $\blacksquare$  &  &  &  &  & $\blacksquare$  & $\blacksquare$  \\ 
         &  93,900  &  5,000  &  4,155  &  32,767  &  20,181  &  3,792  &  & COMPRESSION-3 & $\blacksquare$  & $\blacksquare$  &  & $\blacksquare$  & $\blacksquare$  &  &  &  &  & $\blacksquare$  & $\blacksquare$  \\ 
         &  93,200  &  5,000  &  3,766  &  35,879  &  20,021  &  3,942  &  & COMPRESSION-5 & $\blacksquare$  & $\blacksquare$  &  & $\blacksquare$  & $\blacksquare$  &  &  &  &  & $\blacksquare$  & $\blacksquare$  \\ 
         &  90,800  &  5,000  &  3,585  &  35,731  &  19,517  &  3,986  &  & COMPRESSION-7 & $\blacksquare$  & $\blacksquare$  &  & $\blacksquare$  & $\blacksquare$  &  &  &  &  & $\blacksquare$  & $\blacksquare$  \\ 
         &  90,000  &  5,000  &  3,428  &  35,731  &  19,517  &  3,986  &  & COMPRESSION-9 & $\blacksquare$  & $\blacksquare$  &  & $\blacksquare$  & $\blacksquare$  &  &  &  &  & $\blacksquare$  & $\blacksquare$  \\ 
         \cline{2-20}
         &  1,560  &  5,000  &  1  & 1,049 & 274 &  327  & PNG-FROM-WEB & - & $\blacksquare$  & $\blacksquare$  &  & $\blacksquare$  & $\blacksquare$  &  &  &  &  & $\blacksquare$  & $\blacksquare$  \\ \hline
        PPT &  5,050  &  2,105  &  15  &  20,087  &  2,576  &  3,056  & PPT & - &  &  & $\blacksquare$  & $\blacksquare$  & $\blacksquare$  &  & $\blacksquare$  &  &  &  & $\blacksquare$  \\ 
         &  5,050  &  2,105  &  15  &  20,057  &  2,576  &  3,056  & PPT-NOMAGIC & - &  &  & $\blacksquare$  & $\blacksquare$  & $\blacksquare$  &  & $\blacksquare$  &  &  &  & $\blacksquare$  \\ 
         &  4,820  &  2,105  &  42  &  20,009  &  2,460  &  2,955  & PPT-PASSWORD & - &  &  & $\blacksquare$  & $\blacksquare$  & $\blacksquare$  &  & $\blacksquare$  &  &  &  & $\blacksquare$  \\ \hline
        PPTX &  10,300  &  4,215  &  39  &  118,293  &  2,634  &  4,005  & PPTX & - &  &  & $\blacksquare$  & $\blacksquare$  & $\blacksquare$  &  & $\blacksquare$  &  &  &  & $\blacksquare$  \\ 
         &  10,300  &  4,215  &  39  &  118,293  &  2,634  &  4,005  & PPTX-NOMAGIC & - &  &  & $\blacksquare$  & $\blacksquare$  & $\blacksquare$  &  & $\blacksquare$  &  &  &  & $\blacksquare$  \\ 
         &  10,100  &  4,215  &  46  &  119,254  &  2,576  &  3,993  & PPTX-PASSWORD & - &  &  & $\blacksquare$  & $\blacksquare$  & $\blacksquare$  &  & $\blacksquare$  &  &  &  & $\blacksquare$  \\ \hline
        Powershell &  20  &  5,000  &  1  &  459  &  4  &  12  & PS &  & $\blacksquare$  &  & $\blacksquare$  & $\blacksquare$  & $\blacksquare$  &  &  &  &  &  & $\blacksquare$  \\ \hline
        RANDOM &  17  &  5,000  &  1  &  2  &  1  &  0  & PSEUDO & - &  &  &  &  &  &  &  & $\blacksquare$  &  &  &  \\ 
         &  17  &  5,000  &  1  &  640  &  1  &  9  & PURE & - &  &  &  &  &  &  &  & $\blacksquare$  &  &  &  \\ \hline
        RAR &  17,600  &  5,000  &  6  &  59,321  &  3,793  &  4,750  & RAR & - & $\blacksquare$  &  &  &  &  & $\blacksquare$  &  & $\blacksquare$  & $\blacksquare$  &  &  \\ \hline
        SVG &  122,000  &  5,000  &  4,860  &  48,285  &  26,379  &  5,385  & SVG & - &  & $\blacksquare$  & $\blacksquare$  & $\blacksquare$  & $\blacksquare$  &  &  &  &  & $\blacksquare$  &  \\ 
         &  401  &  5,000  & 1 & 1,049    &  78  & 164   & SVG-FROM-WEB & - &  & $\blacksquare$  & $\blacksquare$  & $\blacksquare$  & $\blacksquare$  &  &  &  &  & $\blacksquare$  &  \\ \hline
        TAR &  20,500  &  5,000  &  20  &  61,542  &  4,421  &  5,339  & TAR & - &  &  &  &  &  &  &  &  &  &  &  \\ \hline
        TIF &  127,000  &  5,000  &  4,149  &  42,664  &  22,900  &  5,019  & TIF & - &  &  &  & $\blacksquare$  &  & $\blacksquare$  &  &  &  &  & $\blacksquare$  \\ 
         &  8,490  &  5,000  &  470  &  3,423  &  1,823  &  466  & TIF-RESIZED & - &  &  &  & $\blacksquare$  &  & $\blacksquare$  &  &  &  &  & $\blacksquare$  \\ \hline
        TXT &  1  &  5,000  &  1  &  231,094  &  5,109  &  28,617  & TXT &  & $\blacksquare$  &  & $\blacksquare$  & $\blacksquare$  & $\blacksquare$  & $\blacksquare$  & $\blacksquare$  &  &  & $\blacksquare$  & $\blacksquare$  \\ \hline
        WEBP &  348  &  5,000  &  1  &  2,4605  &  1,218  &  283  & WEBP-FROM-WEB &  &  & $\blacksquare$  &  &  &  &  &  & $\blacksquare$  &  &  &  \\ 
        \cline{2-20}
         &  5,650  &  5,000  &  344  &  2,460  &  1,214  &  281  & LOSSLESS & COMPRESSION-0 &  & $\blacksquare$  &  &  &  &  &  & $\blacksquare$  &  &  &  \\ 
         &  5,540  &  5,000  &  336  &  2,450  &  1,188  &  280  &  & COMPRESSION-2 &  & $\blacksquare$  &  &  &  &  &  & $\blacksquare$  &  &  &  \\ 
         &  5,540  &  5,000  &  327  &  2,450  &  1,188  &  281  &  & COMPRESSION-4 &  & $\blacksquare$  &  &  &  &  &  & $\blacksquare$  &  &  &  \\ 
         &  5,490  &  5,000  &  327  &  2,450  &  1,179  &  280  &  & COMPRESSION-6 &  & $\blacksquare$  &  &  &  &  &  & $\blacksquare$  &  &  &  \\ 
         \cline{2-20}
         &  555  &  5,000  &  9  &  331  &  114  &  56  & 50\% QUALITY & COMPRESSION-0 &  & $\blacksquare$  &  &  &  &  &  & $\blacksquare$  &  &  &  \\ 
         &  511  &  5,000  &  8  &  333  &  105  &  55  &  & COMPRESSION-2 &  & $\blacksquare$  &  &  &  &  &  & $\blacksquare$  &  &  &  \\ 
         &  451  &  5,000  &  6  &  316  &  93  &  52  &  & COMPRESSION-4 &  & $\blacksquare$  &  &  &  &  &  & $\blacksquare$  &  &  &  \\ 
         &  423  &  5,000  &  6  &  310  &  87  &  50  &  & COMPRESSION-6 &  & $\blacksquare$  &  &  &  &  &  & $\blacksquare$  &  &  &  \\ \hline
        XLS &  2,370  &  5,000  &  5  &  55,651  &  509  &  1,632  & XLS & - &  &  & $\blacksquare$  & $\blacksquare$  & $\blacksquare$  &  & $\blacksquare$  &  & $\blacksquare$  &  & $\blacksquare$  \\ 
         &  2,370  &  5,000  &  5  &  55,651  &  509  &  1,632  & XLS-NOMAGIC & - &  &  & $\blacksquare$  & $\blacksquare$  & $\blacksquare$  &  & $\blacksquare$  &  & $\blacksquare$  &  & $\blacksquare$  \\ 
         &  2,420  &  5,000  &  20  &  55,620  &  518  &  1,646  & XLS-PASSWORD & - &  &  & $\blacksquare$  & $\blacksquare$  & $\blacksquare$  &  & $\blacksquare$  &  & $\blacksquare$  &  & $\blacksquare$  \\ \hline
        XLSX &  2,660  &  5,000  &  8  &  93,930  &  570  &  2,769  & XLSX & - &  &  & $\blacksquare$  & $\blacksquare$  & $\blacksquare$  &  & $\blacksquare$  &  & $\blacksquare$  &  & $\blacksquare$  \\ 
         &  2,660  &  5,000  &  8  &  93,930  &  570  &  2,769  & XLSX-NOMAGIC & - &  &  & $\blacksquare$  & $\blacksquare$  & $\blacksquare$  &  & $\blacksquare$  &  & $\blacksquare$  &  & $\blacksquare$  \\ 
         &  2,700  &  5,000  &  15  &  96,190  &  578  &  2,815  & XLSX-PASSWORD & - &  &  & $\blacksquare$  & $\blacksquare$  & $\blacksquare$  &  & $\blacksquare$  &  & $\blacksquare$  &  & $\blacksquare$  \\ \hline
        XML &  320  &  5,000  &  1  &  16,920  &  75  &  708  & XML &  & $\blacksquare$  &  & $\blacksquare$  & $\blacksquare$  & $\blacksquare$  &  & $\blacksquare$  &  & $\blacksquare$  & $\blacksquare$  & $\blacksquare$  \\ \hline
        ZIP &  18,200  &  5,000  &  7  &  59,865  &  3,912  &  4,852  & BZIP2 & - & $\blacksquare$  &  &  &  & $\blacksquare$  & $\blacksquare$  & $\blacksquare$  & $\blacksquare$  &  &  & $\blacksquare$  \\ 
         &  18,200  &  5,000  &  6  &  59,762  &  3,907  &  4,855  & DEFLATE & - & $\blacksquare$  &  &  &  & $\blacksquare$  & $\blacksquare$  & $\blacksquare$  & $\blacksquare$  &  &  & $\blacksquare$  \\ 
         &  17,600  &  5,000  &  6  &  59,352  &  3,790  &  4,740  & LZMA & - & $\blacksquare$  &  &  &  & $\blacksquare$  & $\blacksquare$  & $\blacksquare$  & $\blacksquare$  &  &  & $\blacksquare$  \\ 
         &  18,000  &  5,000  &  6  &  60,660  &  3,878  &  4,831  & PPMD & - & $\blacksquare$  &  &  &  & $\blacksquare$  & $\blacksquare$  & $\blacksquare$  & $\blacksquare$  &  &  & $\blacksquare$  \\ 
         &  18,100  &  5,000  &  6  &  59,695  &  3,896  &  4,845  & HIGH-COMPRESSION & - & $\blacksquare$  &  &  &  & $\blacksquare$  & $\blacksquare$  & $\blacksquare$  & $\blacksquare$  &  &  & $\blacksquare$  \\ 
         &  18,200  &  5,000  &  7  &  59,762  &  3,907  &  4,855  & ENCRYPTED & - & $\blacksquare$  &  &  &  & $\blacksquare$  & $\blacksquare$  & $\blacksquare$  & $\blacksquare$  &  &  & $\blacksquare$  \\ \hline
        ZLIB &  18,200  &  5,000  &  6  &  59,762  &  3,912  &  4,860  & ZLIB & - & $\blacksquare$  &  &  &  &  &  &  & $\blacksquare$  &  &  &  \\ \hline
        7Zip &  18,200  &  5,000  &  7  &  59,864  &  3,915  &  4,852  & BZIP2 & - & $\blacksquare$  &  &  &  &  & $\blacksquare$  &  & $\blacksquare$  &  &  &  \\ 
         &  17,600  &  5,000  &  6  &  59,351  &  3,785  &  4,738  & LZMA & - & $\blacksquare$  &  &  &  &  & $\blacksquare$  &  & $\blacksquare$  &  &  &  \\ 
         &  17,500  &  5,000  &  6  &  59,015  &  3,776  &  4,724  & LZMA2 & - & $\blacksquare$  &  &  &  &  & $\blacksquare$  &  & $\blacksquare$  &  &  &  \\ 
         &  18,000  &  5,000  &  6  &  60,626  &  3,881  &  4,830  & PPMD & - & $\blacksquare$  &  &  &  &  & $\blacksquare$  &  & $\blacksquare$  &  &  &  \\ 
         &  17,500  &  5,000  &  6  &  58,790  &  3,775  &  4,721  & HIGH-COMPRESSION & - & $\blacksquare$  &  &  &  &  & $\blacksquare$  &  & $\blacksquare$  &  &  &  \\ 
         &  17,500  &  5,000  &  6  &  59,016  &  3,776  &  4,724  & ENCRYPTED & - & $\blacksquare$  &  &  &  &  & $\blacksquare$  &  & $\blacksquare$  &  &  &  \\ \hline
        EXTRA &  4,410  &  5,000  &  7  &  32,156  &  945  &  2,086  & RANSOMWARE & NOTPETYA &  &  &  &  &  &  &  & $\blacksquare$  &  &  &  \\ 
         &  3,740  &  5,000  &  1  &  32,156  &  802  &  1,951  &  & SODINOKIBI &  &  &  &  &  &  &  & $\blacksquare$  &  &  &  \\ 
         &  3,730  &  5,000  &  1  &  32,156  &  801  &  1,951  &  & MAZE &  &  &  &  &  &  &  & $\blacksquare$  &  &  &  \\ 
         &  4,190  &  5,000  &  1  &  44,381  &  954  &  2,283  &  & PHOBOS &  &  &  &  &  &  &  & $\blacksquare$  &  &  &  \\ 
         &  3,740  &  5,000  &  1  &  32,156  &  802  &  1,951  &  & NETWALKER &  &  &  &  &  &  &  & $\blacksquare$  &  &  &  \\ 
         &  3,902  &  5,000  &  1  &  32,257  &  804  &  1,951  &  & DHARMA &  &  &  &  &  &  &  & $\blacksquare$  &  &  &  \\ 
         &  3,994  &  5,000  &  1  &  31,580  &  802  &  1,951  &  & RYUK &  &  &  &  &  &  &  &  &  &  &  \\ 
         \arrayrulecolor{black}
         \bottomrule
    \end{tabular}
    \endgroup

\label{tab:data-set-details}
\end{table*}

\begin{figure*}
\centering
  \includegraphics[width=0.9\textwidth]{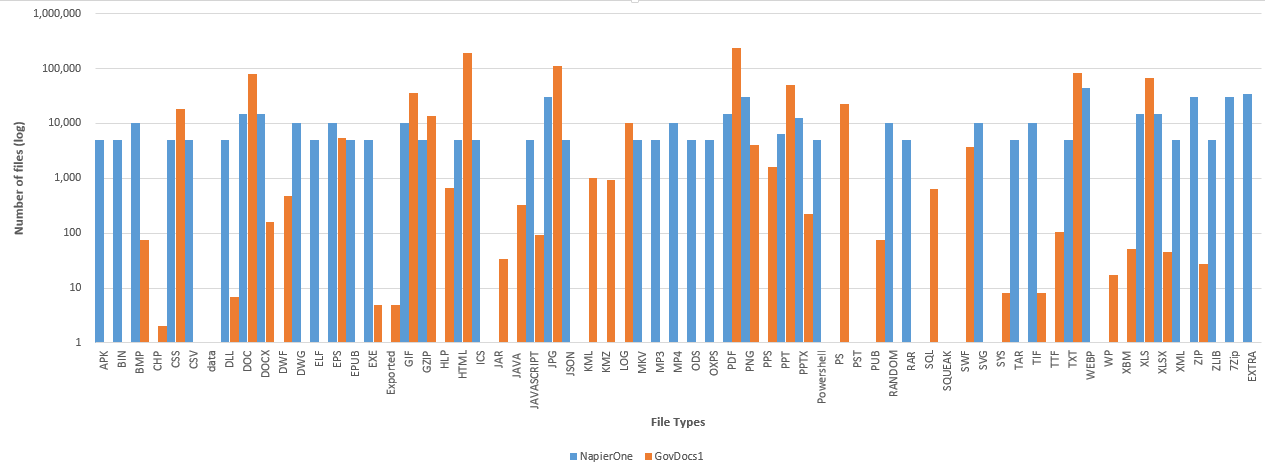}
  \caption{Dataset Comparison.}
    \label{fig:dataset-comparison}
\end{figure*}

\subsection{Data Set Comparison}
\noindent
When comparing the NapierOne and Govdocs1 data sets, several differences become apparent. While both data sets contain approximately 40 different distinct file types, the majority of the files in the Govdocs1 dataset represent only a few of these file types. For example, half of the dataset consists of only three file types (HTML, JPG and PDF). Whereas it can be seen from Figure{~\ref{fig:dataset-comparison}} that the NapierOne data set file distribution is considerably more even across all the represented files. For nine of the file types present in the Govdocs1 data set, there exists less than 10 example files, whereas in the NapierOne dataset most file types have at least 5,000 example files.
The content of the Govdocs1 data set was determined by the files found using random searches of the \emph{.gov} domain and may not be representative of other systems or file type popularity in general. The content of the NapierOne data set was determined based on the research performed in to file usage. This real-world file usage information being gathered from more than 11 separate sources. The collected usage statistics were then collated, aggregated and ranked according to frequency and then used to determine the constituents of the NapierOne data set. 

There also exist many more examples of modern file formats such as DOCS, PPTX and XLSX in the NapierOne dataset than are present in the Govdocs1 dataset. Finally, there are many more examples of files with high entropy such as archive, compressed, random data and encrypted files in the NapierOne dataset. These types of files are useful when researching, for example, ransomware detection systems. Actual specific data set information can be found in Tables{~\ref{tab:govdocstypes}} and {~\ref{tab:data-set-details}}.

\subsection{Access}
\noindent As long as the attribution and license stipulations are followed, access to this \emph{NapierOne}~\cite{Davies2021} data set for research purposes is free and unrestricted. For access to the data set, visit \textbf{www.napierone.com} or alternatively contact the authors of this paper.

It can be seen from Table~\ref{tab:data-set-details} that some data subsets are extremely large, with the total uncompressed size of the entire data set approaching 2TB. Due to its size, the data set has been broken down into smaller parts allowing the possibility to access only portions of the data set.
The final data set is broken down as follows.
\begin{itemize}
    \item Each data subset is available in three separate archives.
    \begin{itemize}
        \item \textbf{Tiny} This archive contains 100 examples of the specific file type. The actual files contained within this archive are the files of this type that have sequence numbers between 0001 and 0100.
        \item \textbf{Small} This archive contains 1,000 examples of the specific file type. The actual files contained within this archive are the files of this type that have sequence numbers between 0001 and 1,000.
        \item \textbf{Total} This archive contains all the examples of the specific file type present in the \emph{NapierOne} data set.
    \end{itemize}
    \item These individual archives have also been combined into three larger archives as described below:
    \begin{itemize}
        \item \textbf{NapierOne-Tiny} This archive contains a copy of all the \emph{Tiny} archives mentioned in the bullet point above.
        \item \textbf{NapierOne-Small} This archive contains a copy of all the \emph{Small} archives mentioned in the bullet point above.
        \item \textbf{NapierOne-Total} This archive contains a copy of all the \emph{Total} archives mentioned in the bullet point above. This archive is very large as it contains all the data from the \emph{NapierOne} data set.
    \end{itemize}
\end{itemize}

Each data set archive is accompanied by documentation providing the user with a detailed description of that particular archive including attribution and licensing instructions. Using this structured approach, researchers will be able to tune, what and how much they download, to their specific requirements. Password protected versions of document and archive files all use the same password, \emph{napierone}. 

When accessing the website some simple registration may be required. This was implemented to allow the authors to gather some basic usage information regarding the data set. Information gathered during the registration will be handled in accordance with GDPR regulations~\cite{Union2017}. The anonymised information gathered during the registration will allow the authors to monitor the use, application and adoption of the data set.

\section{Conclusion}
\label{cha:conclusion}
\noindent Inspired by the work performed by Garfinkel et. al.\cite{Garfinkel2009}, this paper describes the design, development and distribution of the \emph{NapierOne}~\cite{Davies2021} mixed file data set. This data set creation was guided by the approaches and techniques used in the development of the Govdocs1~\cite{Garfinkel2009} and attempts to complement the original data set with examples of file types that have appeared, or increased in popularity, since its original release. There is no duplication of data between to original Govdocs1 and \emph{NapierOne} data sets.

The data set attempts to address a reoccurring theme in the literature where it has been shown that there exists a lack of useful data sets available for researchers and how the default behaviour of many researchers is to generate their own data sets and not distribute them. A consequence of this behaviour is that research validation and reproducibility is impacted. Many papers highlight the opinion that increased accessibility to high-quality data sets would undoubtedly improve the speed and direction of research and in so doing facilitate high-quality research, that is easier to peer review, producing more robust solutions to the issues being investigated.

The paper describes how the specific file types within this data set were selected, how examples were sourced and how researchers are able to gain free, unlimited access to the data. Actions performed to ensure that the data contained within the data set is of the highest quality are also described. The breadth and depth of the \emph{NapierOne} data set should also remove the need for other researchers to complement the data set with additional files as has previously been the case~\cite{Nguyen2014,Penrose2013,Pont2020}.


The \emph{NapierOne} data set is an ongoing project and the authors intend to employ the developed data set as part of their research in to ransomware detection techniques. Researchers are strongly encouraged to leverage the data set in their own research. The authors are also open to any comments or discussion concerning any aspect of this data set.

\section{Postscript}
\label{cha:Postscript}
\noindent The authors consider NapierOne data set to be a starting point data set and act as a foundation for the creation of a much larger corpus. 

The authors have attempted to validate the content of the data set to the best of their ability. This included visual inspection of image files, submission to malware scanning services, local virus checking as well as programmatically checking the format of each file to confirm that it complies with that specific format and can be opened by a program designed to handle that format.

It is hoped that over time other researchers will provide additional data sets that can be incorporated into NapierOne allowing it to grow and develop in the spirit of Open Source.

\printcredits

\bibliographystyle{cas-model2-names}

\bibliography{bibliography}

\end{document}